\documentclass{article}
\usepackage{amssymb}
\usepackage{amsmath}
\usepackage{epsfig}
\renewcommand{\theequation}{\thesection.\arabic{equation}}
\newcommand\supsetplus{\leavevmode\raise.11em\hbox{\tiny +}
  {\kern-.49em\hbox{$\supset$}}}
\newcommand\jedop{\leavevmode\raise.11em\hbox{\scriptsize 1}
  {\kern-.95em\hbox{$\bigcirc \, $}}}
\newcommand\dwaop{\leavevmode\raise.11em\hbox{\scriptsize 2}
  {\kern-.98em\hbox{$\bigcirc \, $}}}
\newcommand\trzop{\leavevmode\raise.11em\hbox{\scriptsize 3}
  {\kern-.98em\hbox{$\bigcirc \, $}}}
\newcommand\cztop{\leavevmode\raise.11em\hbox{\scriptsize 4}
  {\kern-.95em\hbox{$\bigcirc \, $}}}
\newcommand\bcoper{\vartriangleright{\kern-.55em\hbox{$\blacktriangleleft \,
$}}}
\begin{document}

\title{Quantum Deformations of Space--Time SUSY  and
 Noncommutative
       Superfield Theory}
                        
\author{P. Kosi\'{n}ski 
     \\ 
Institute of Physics,
           University of L\'{o}d\'{z}, 
\\ ul. Pomorska 149/53 90--236
           L\'{o}d\'{z},   Poland ]
\\ \\
J. Lukierski
 \\
email:lukier@ift.uni.wroc.pl
                            \\
Institute of Theoretical Physics, University of Wroclaw 
\\ pl. M. Borna 9,
           50-205 Wroc\l aw, Poland 
                                    \\ \\
P. Ma\'{s}lanka \\
 Institute of
        Physics, University of L\'{o}d\'{z}, \\ ul. Pomorska 149/53
        90--236 L\'{o}d\'{z}, Poland }

\date{}
\maketitle 

\begin{abstract}
We review shortly present status of quantum deformations of
Poincar\'{e} and conformal supersymmetries. After recalling the
$\kappa$--deformation of $\hbox{D=4}$
 Poincar\'{e} supersymmetries we describe the corresponding star
 product multiplication for chiral
 superfields.
 In order to describe the deformation of
  chiral vertices in momentum space
 the integration formula over $\kappa$--deformed
 chiral superspace is proposed.
\end{abstract}

    \section{Introduction}
The noncommutative space--time coordinates were introduced as describing
 algebraically
 the quantum gravity corrections to commutative flat
 (Minkowski) background (see
 e.g.
\cite{luk-dop,luk-gar}) as well as the modification of $D$--brane
 coordinates in the presence of
 external background tensor fields (e.g. $B_{\mu\nu}$
 in $D=10$ string theory;
 see
\cite{luk-chon}--\cite{luk-mad}). We know well that both gravity and string  theory have
 better properties (e.g. less divergent quantum perturbative expansions)
 after  their  supersymmetrization. It appears  therefore reasonable,
 if not compelling,  to  consider the supersymmetric
  extensions of the noncommutative
 framework.

The generic relation for the noncommutative space--time generators
$\widehat{x}_{\mu}$

\begin{eqnarray}
\label{luk-lukjedcc}
[\widehat{x}_{\mu}, \widehat{x}_{\nu}]=
i\Theta_{\mu\nu}{(\widehat{x})}= i\left(\Theta_{\mu\nu} +
\Theta_{\mu\nu}^{\rho} \widehat{x}_{\rho} + \ldots \right)
\end{eqnarray}
has been usually considered for constant value of the commutator
 (\ref{luk-lukjedcc}), i.e. for $\Theta_{\mu\nu}(\widehat{x}) = \Theta_{\mu\nu}$.
In such a case the multiplication of the fields
$\phi_{k}{(\widehat{x})}$ depending on the noncommutative (Minkowski)
space--time  coordinates can be represented by noncommutative
Moyal $\ast$--product of classical fields
 $\phi_{k}{(x)}$ on standard Minkowski space
\begin{eqnarray}
\label{luk-dwa}
  \phi_{k}(\widehat{x}) \phi_{l} (\widehat{x} )\longleftrightarrow
  \phi_{k}{(x)}\ast \phi_{l}{(x)} = \phi_{k}(y)
  e^{{i\over 2}
  \Theta^{\mu\nu}
   {\partial\over \partial y_{\mu} }
  {\partial\over  y_{\nu} }} \phi_{l}(z)|_{x=y}
\end{eqnarray}
It appears that the relation
 (\ref{luk-lukjedcc}) with constant $\Theta_{\mu\nu}$
 can be consistently supersymmetrized
 (see e.g. \cite{luk-chu}--\cite{luk-dop2}) by
 supplementing the standard relations for the odd Grassmann
 superspace coordinates (further  we choose $D=4$ \, $N=1$\,  SUSY
  and $\alpha,
 \beta=1,2$).
\begin{eqnarray}
\label{luk-trz}
  \{\theta_{\alpha},\theta_{\beta}\}
  = \{\theta_{\alpha},\overline{\theta}_{\dot{\beta}}
  \}
   = \{\overline{\theta}_{\dot{\alpha}},
  \overline{\theta}_{\dot{\beta}}\} =0 \qquad
  [\widehat{x}_{\mu}, \theta_{\alpha} ]
  =[\widehat{x}_{\mu}, \overline{\theta}_{\dot{\alpha}}]
  = 0
\end{eqnarray}
Such a choice of superspace coordinates ($\widehat{x}_{\mu},
\theta_{\alpha},  \theta_{\dot{\alpha}}$) implies that the
supersymmetry transformations remain classical:
\begin{eqnarray}
\label{luk-czt}
  \widehat{x}^{\prime}_{\mu}=
  \widehat{x}_{\mu}- i\left( \overline{\epsilon}\sigma_{k}
\theta_{\alpha}-
  \overline{\theta}\sigma_{k}\epsilon \right)
  \cr\cr
  \theta^{\prime_{\alpha}}=\theta_{\alpha}+ \epsilon_{\alpha}
  \qquad
  \overline{\theta}^{\prime_{\dot{\alpha}}}=
  \overline{\theta_{\dot{\alpha}}}+\overline{\epsilon_{\dot{\alpha}}}
\end{eqnarray}
i.e. the covariance requirements  of deformed superspace formalism
do not require the deformation of classical Poincar\'{e}
supersymmetries\footnote{It should be stressted, however, that
 the introduction of constant tensor $\Theta_{\mu\nu}$ in
 (\ref{luk-lukjedcc})
 leads to breaking ($O(3,1)      \to O(2)\times  O(1,1))$) of
 $D=4$ Lorentz symmetry. The way out is to consider  $\Theta_{\mu\nu}$
  as a constant field, with generator of Lorentz subalgebra
  containing  contribution which rotates the  $\Theta_{\mu\nu}$
  components  (see e.g.
\cite{luk-dopp}). The relation
 (\ref{luk-lukjedcc})  can be made
  covariant only for $D=2$ ($\Theta_{\mu\nu}\equiv \epsilon_{\mu\nu}$
  for $D=2$); for 2+1 Euclidean case see \cite{luk-lukx}}.

  Our aim here is  to consider the case  when the  standard Poincar\'{e}
  supersymmetries can not be preserved. For this purpose     we
  shall consider the  case with linear Lie--algebraic commutator
 (\ref{luk-lukjedcc}). Its supersymmetrization leads to the deformed superspace
  coordinates $\widehat{z}_{A}=(
  \widehat{x}_{\mu},\widehat{\theta}_{\alpha},
  \widehat{\overline{\theta}}_{\dot{\beta}})$ satisfying Lie
  superalgebra  relation:
\begin{eqnarray}
\label{luk-pie}
  [\widehat{z}_{A},\widehat{z}_{B}] = i\Theta_{AB}^{C}\, \widehat{z}_{C}
\end{eqnarray}
where $\Theta_{AB}^{C}$ satisfies graded Jacobi identity:
\begin{eqnarray}
\label{luk-sze}
\Theta_{AB}^{\ D}\, \Theta_{CD}^{\ E} + \mbox{graded cycl.}
 \ (A,B,C) \qquad = 0
\end{eqnarray}
It appears that in such a case for some choices of the ``structure
constants" $\Theta_{AB}^{\ C}$ one can find  the deformed quantum
$D=4$ Poincar\'{e} supergroup, which provide the relations
(\ref{luk-pie})
as describing the deformed translations and deformed
supertranslations.

The plan of the paper is following: In Sect. 2  we shall briefly
review the considered in literature quantum deformations  of
Poincar\'{e} and conformal supersymmetries. The list of these
deformations written in explicite form as    Hopf algebras is
quite short, and only the knowledge of large class of classical
$r$--matrices shows that many quantum deformations should be still
discovered. As the      only nontrivial
 quantum deformation of $D=4$
supersymmetry     given in the literature is the so--called
$\kappa$--deformation, obtained in 1993
\cite{luk-luk1}--\cite{luk-kos2}.

In Sect. 3 we consider the Fourier supertransform of superfields in
 classical (undeformed) and $\kappa$--deformed form. We present also the
integration formula over $\kappa$--deformed superspace, which provides
 the description in supermomentum space leading to the
$\kappa$--deformed Feynmann superdiagrams.

In Sect. 4 we consider the $\kappa$--deformed superfield theory in
  chiral superspace. We introduce the $\ast$--product
multiplication of $\kappa$--deformed  superfields. It appears that
there  are two distinguished $\ast$--products, which both can be
written in closed form: one described by standard supersymmetric
extenion of CBA  formula and other physical, providing the
addition of fourmomenta and Grassmann momenta in terms of the
coproduct formulae. In such a way we obtain the supersymmetric
extension of two $\ast$--products, considered recently in
\cite{luk-kos3}.

In Sect. 5 we shall  present some remarks   and  general diagram describing
 the deformation scheme of superfield theory.

\section{Quantum Deformations of Space--Time Supersymmetries}
\setcounter{equation}{0}

 There are two  basic space--time
symmetries in D dimensions:

- Conformal symmetries $O(D,2)$, having another interpretation as
anti--de--Sitter symmetries in $D+1$ dimensions

-  Poincar\'{e} symmetries $T^{D-1,1}  \supsetplus O(D-1,1)$.

i) Quantum deformations of conformal supersymmetries.

The conformal symmetries can be supersymmetrized  without
introducing tensorial central charges in $D=1,2,3,4$ and 6. One
gets:

\begin{eqnarray}
  D=1 & : & O(2,1) \longrightarrow OSp(N;2|R)  \quad \mbox{or}
  \quad SU(1,1:N) \cr\cr
  D=2 &: & O(2,2)=O(1,2)\otimes O(1,2) \longrightarrow
  OSp(M;2|R) \otimes OSp(N;2| R)  \mbox{}
    \cr\cr
  D=3 & & O(3,2) \longrightarrow OSp(N;4|R) \cr\cr
D=4   && O(4,2)  \longrightarrow SU (2,2;N) \cr\cr
  D=6 && O(6,2) \longrightarrow  U_{\alpha} U(4;N|H)\nonumber
\end{eqnarray}

All  conformal supersymmetries  listed above are described by
simple Lie superalgebras. It is well--known that for every simple
Lie superalgebra one can introduce the $q$--deformed
Cartan--Chevaley basis describing quantum (Hopf--algebraic)
Drinfeld--Jimbo deformation
 \cite{luk-chai,luk-khoro}. These $q$--deformed relations
have been explicitly written in physical basis of conformal
superalgebra in different dimensions (see e.g.
\cite{luk-luk3}--\cite{luk-??3}). It is easy  to see that the
deformation parameter $q$ appears as dimensionless.

It follows, however, that there is another class of deformations
of conformal and superconformal symmetries, with dimensionfull
parameter $\kappa$, playing  the role of geometric fundamental
mass. For $D=1$    one can show that the Jordanian deformation of
$SL(2;R)\simeq O(2,1)$ describes the $\kappa$--deformation of
$D=1$ conformal algebra
 \cite{luk-luk4}. This result can be extended
supersymmetrically, with the following classical
$\widehat{r}$--matrix  describing Jordanian deformation
$U_{\kappa}(OSp(1;2|R)$
\cite{luk-jusz}
\begin{eqnarray}
\label{luk-dwje}
  r = {1\over \kappa}\, h \wedge e
\qquad  & \stackrel{SUSY}{\Longrightarrow}
 &  r ={1\over \kappa} \left(  h\wedge e + Q^{+} \wedge Q^{+}\right)
  \cr\cr
  \mbox{Jordanian deformation} && \mbox{Jordanian deformation}
  \cr
  \mbox{of}\  Sp(2;R)\simeq O(2;1;R) &&  \mbox{of} \  OSp(1,2;R)
  \cr
   (D=1 \quad \mbox{conformal} )
   && (D=1 \quad  \mbox{superconformal} )
\end{eqnarray}

The $OSp(1;2;R)$ Jordanian classical $\widehat{r}$--matrix can be
quantized by the twist method. Semi--closed form for the twist
function has been obtained in
 \cite{luk-cele}.

It appears that one can extend the Jordanian deformations of $D=1$
conformal algebra to $D > 1$;  for $D=3$ and $D=4$ the extended
Jordanian classical $r$--matrices were given in
\cite{luk-luk4}. It should be also mentioned that  the generalized
Jordanian deformation
 of $D=3$ conformal $O(3,2)$ algebra   has been obtained in full
Hopf--algebraic form
 \cite{luk-herr}. The extension of
Jordanian deformation of $OSp(1,2;R)$ for $D> 1$ superconformal algebras
  is not known even
in its infinitesimal form given by classical $r$--matrices.

ii) Quantum deformations of Poincar\'{e} supersymmetries.

Contrary to DJ  scheme for simple   Lie (super)algebras it does
not exist a systematic way of obtaining quantum deformations of
non--semisimple Lie (super)algebras. A natural framework for the
description of deformed semi--direct products, like quantum
Poincar\'{e} algebra, are the noncocommutative bicrossproduct Hopf
algebras (see e.g.
\cite{luk-moy}). It appears
however that
 in the literature
 it  has not been formulated  any effective scheme
describing these quantum bicrossproducts.

One explicite example of quantum deformation of $D=4$ Poincar\'{e}
superalgebra and its dual $D=4$ Poincar\'{e} group in form of
graded bicrossproduct Hopf algebra was given in
\cite{luk-kos2}.
 By means of
quantum contraction of $q$--deformed $N=1$ anti--de--Sitter
superalgebra $U_{q}$ ($OSp(1|4))$ there was obtained in
\cite{luk-luk1} the $\kappa$--deformed $D=4$ Poincar\'{e}
subalgebra $U_{\kappa}({\cal P}_{4;1})$. Subsequently by nonlinear
change of generators the quantum superalgebra $U_{\kappa}({\cal
P}_{4;1})$ was written in chiral bicrossproduct  basis
\cite{luk-kos1}. The
$\kappa$--deformed Poincar\'{e} subalgebra is given by the
deformation of the following   graded cross--product
\footnote{In 
\protect\cite{luk-kos1} for the crossproduct
formula describing $D=4$  superPoincar\'{e} algebra the
following notation was used: $p_{4;1} = O(1,3;2)\ltimes T_{4,2}$.
In the notation
 (\ref{luk-dwdw})  proposed in present paper    the extension
of Lorentz algebra by odd  generators is described more
accurately.}
\begin{eqnarray}
\label{luk-dwdw}
  p_{4;1} = \left( SL(2;C)\oplus (\overline{SL(2;C)}
  \supsetplus \, \overline{T}_{0;2}\right)\ltimes
  T_{4;2}
\end{eqnarray}
where the generators of $SL(2;C)$ are given by two--spinor
generators $M_{\alpha\beta} = {1\over
8}(\sigma^{\mu\nu})_{\alpha\beta}M_{\mu\nu}$, the generators of
 $(\overline{SL(2;C)}$ by $M_{\dot{\alpha}\dot{\beta}}=
 M_{\alpha\beta}^{\ast} = {1\over 8}
 \sigma^{\mu\nu}_{\dot{\alpha}\dot{\beta}}
 M_{\mu\nu}$,  $\overline{T}_{0;2}$  describes two antichiral
 supercharges  $\overline{Q}_{\dot{\alpha}}$, and $T_{4;2}$ the
 graded Abelian superalgebra

\begin{eqnarray}
\label{luk-dwtrz}
T_{4;2} :\qquad [  P_{\mu}, P_{\nu}] = [P_{\mu},Q_{\alpha}] =
\{Q_{\alpha}, Q_{\beta} \} = 0
\end{eqnarray}
The  relations
 (\ref{luk-dwtrz})  describe the algebra of   generators of
translations and supertranslations in chiral superspace. The
algebra $(SL(2;c)\oplus(\overline{SL(2;c)}   \supsetplus \,
\overline{T}_{0;2})$
 has the form

\renewcommand{\theequation}{\thesection.4\alph{equation}}
\setcounter{equation}{0}
\begin{eqnarray}
\label{luk-dwcztery}
sl(2;c): \qquad  [M_{\alpha\beta},M_{\gamma\delta}]
& = &  \epsilon_{\alpha\gamma}M_{\beta\delta} -
\epsilon_{\beta\gamma}M_{\alpha\delta}
 \\ \nonumber
&& + \  c_{\beta\delta}
M_{\alpha\gamma} - \epsilon_{\alpha\delta}M_{\beta\gamma}
\\ \nonumber
\overline{sl(2;c)}  \supsetplus \, \overline{T}_{0;2}:
\qquad
[M_{\dot{\alpha}\dot{\beta}},M_{\dot{\gamma}\dot{\delta}}]
&  =  &
\epsilon_{\dot{\alpha}\dot{\gamma}}M_{\dot{\beta}\dot{\delta}}
-\epsilon_{\dot{\beta}\dot{\gamma}} M_{\dot{\alpha}\dot{\delta}}
\\ \nonumber
 &&  + \
\epsilon_{\dot{\beta}\dot{\delta}}M_{\dot{\alpha}\dot{\gamma}} -
\epsilon_{\dot{\alpha}\dot{\delta}}M_{\dot{\beta}\dot{\gamma}}
\\  \nonumber
 [M_{\dot{\alpha}\dot{\beta}},Q_{\dot{\gamma}}]   &=  &
\epsilon_{\dot{\alpha\gamma}}Q_{\dot{\beta}}
-\epsilon_{\dot{\beta}\dot{\gamma}}Q_{\dot{\alpha}}
 \\
\{
Q_{\dot{\alpha}},Q_{\dot{\beta}}\} & = &  0
\label{luk-dwczttb}
\end{eqnarray}

\renewcommand{\theequation}{\thesection.\arabic{equation}}
\setcounter{equation}{4}

It should be observed that in the cross-product
 (\ref{luk-dwdw}) the basic
supersymmetry algebra $\{Q_{\alpha},Q_{\dot{\beta}}
\}=2(\sigma^{\mu} p_{\mu} )_{\alpha\dot{\beta }}$ is the one belonging
to the cross--relations.

The  $\kappa$--deformed bicrossproduct is given by the formula
\begin{eqnarray}
\label{luk-dwpi}
  U_{\kappa}(p_{4;2}) = (SL(2;c)\oplus \overline{SL(2;c)}
   \supsetplus \, \overline{T}_{0;2})\bcoper T^{\kappa}_{4;2}
\end{eqnarray}
The relations
 (\ref{luk-dwtrz}) and
 (\ref{luk-dwcztery})--(\ref{luk-dwczttb})  remain valid but $
T^{\kappa}_{4;2}$ describes now the  Hopf algebra with deformed
coproducts:
\begin{eqnarray}
\label{luk-dwszescc}
\Delta P_{0} & =& P_{0} \otimes 1 + 1 \otimes P_{0} \cr \Delta
P_{i} & =& P_{i} \otimes e^{-{P_{0}\over \kappa}} + 1 \otimes
P_{i} \cr
 \Delta Q_{\alpha} & =& Q_{\alpha} \otimes
  e^{-{P_{0}\over 2 \kappa}} + 1 \otimes Q_{\alpha}
\end{eqnarray}

The cross--relations are the following $(M_{i}= {1\over 2}
\epsilon_{ijk}M_{jk}, N_{i}=M_{i0})$:

\begin{eqnarray}
\label{luk-dwsiedem}
[M_{i},P_{j}] &=& i\epsilon_{ijk}P_{k} \qquad  [M_{i},P_{0}] =0
\cr\cr
 [  N_{i},P_{j} ] &= & i\delta_{ij}[ {\kappa \over 2} (1-
e^{-{2P_{0}\over\kappa} } + {1\over 2\kappa }
\stackrel{\rightharpoonup}{P} ^{\ 2}) + {1\over \kappa} P_{i}P_{j}]
\cr\cr
[ N_{i},P_{0}] &=& iP_{i}
\end{eqnarray}
and

\begin{eqnarray}
[M_{i},Q_{\alpha}     ] &=& - {1\over 2} (\sigma_{i})_{\alpha} ^{\
\beta} Q_{\beta} \cr\cr [ N_{i},Q_{\alpha} ] & =& {1\over 2} i \,
 e^{- {P_{0} \over \kappa}} (\sigma_{i})_{\alpha}^{\ \beta}
 Q_{\beta} + {1\over 2\kappa} \,
 \epsilon_{ijk}P_{j}(\sigma_{k})_{\alpha}^{\ \beta} Q_{\beta}
 \cr\cr
 \{Q_{\alpha},Q_{\dot{\beta}} \} &= & 4\kappa
 \delta_{\alpha\dot{\beta}} \sinh {P_{0}\over 2\kappa} - 2
 e^{P_{0}\over 2\kappa}     p_{i}(\sigma_{i})_{\alpha \dot{\beta}}
\end{eqnarray}
The notion of bicrossproduct
  (\ref{luk-dwpi}) implies also the modification
of primitive coproducts  for $SL(2;c)\oplus \overline{SL(2;c)}
  \supsetplus \, \overline{T}_{0;2}$ generators. One   gets:
 \begin{eqnarray}
 \Delta M_{i} & = & M_{i}      \otimes 1  + 1 \otimes M_{i}
\cr\cr \Delta N_{i} &= & N_{l} \otimes 1 + e^{- {P_{0}\over
\kappa}} \otimes N_{i} + {1\over \kappa} \epsilon_{ijk}    \,
P_{j}\otimes M_{k}\cr && - \ {i\over 4\kappa}
(\sigma_{i})_{\alpha\dot{\beta}} \, Q_{\alpha} \otimes
e^{P_{0}\over \kappa} Q_{\dot{\beta}} \cr\cr \Delta Q_{j} & = &
Q_{\dot{\alpha}} \otimes 1 + e^{P_{0}\over 2 \kappa} \otimes
Q_{\dot{\alpha}}
\end{eqnarray}
It appears that the classical $N=1$  $D=4$  Poincar\'{e}
superalgebra  can be  put as well  in the form
\begin{eqnarray}
\label{luk-dwdzi}
  p_{4;1} = \left( SL(2;c)
  \supsetplus \, {T}_{0;2}\right)
\oplus (\overline{SL(2;c)}
  \ltimes
  \overline{T}_{4;2}
\end{eqnarray}
where $\overline{T}_{4;2}^{\ 0}$ describe the translation and
supertranslation  generators ($P_{\kappa},Q_{\dot{\alpha}}$).
 Subsequently the
$\kappa$--deformation of $D=4$ $N=1$ Poincar\'{e} superalgebra can
be obtained by deforming
 (\ref{luk-dwdzi}) into graded  bicrossproduct Hopf
superalgebra

\begin{eqnarray}
\label{luk-dwdwa}
  U_{\kappa}(p_{4;1}) = \left( SL(2;c)
    \supsetplus \, {T}_{0;2}\right)
  \oplus (\overline{SL(2;c)}
  \bcoper
  \overline{T}_{4;2}^{\  \kappa}
\end{eqnarray}

In order to describe the $\kappa$--deformed chiral superspace one
should consider the Hopf superalgebra $\widetilde{T}_{4;2}^{\  \kappa}$
obtained by dualization of the relations
      (\ref{luk-dwtrz}) and
 (\ref{luk-dwszescc}), and
describing by functions $C(\widehat{z}_{A})$ on $\kappa$--deformed chiral
superspace $\widehat{z}_{A} = (\widehat{z}_{\mu}, 
\widehat{\theta}_{\alpha})$, where
$\widehat{z}_{\mu}$ denotes the complex space--time coordinates. One obtains
the following set of relations:

\renewcommand{\theequation}{\thesection.12\alph{equation}}

\setcounter{equation}{0}

\begin{eqnarray}
\label{luk-dwanaxxa}
&&[\widehat{z}_{0}, \widehat{z}_{i}]  = {i\over \kappa}
\widehat{z}_{i} \qquad \qquad
[\widehat{z}_{i},\widehat{z}_{j}] = 0
\cr\cr && [\widehat{z}_{0}, \widehat{\theta}_{\alpha} ] =
{i\over 2\kappa} \widehat{\theta}_{\alpha} \qquad
[\widehat{z}_{i},\widehat{\theta}_{\alpha} ] = 0
\cr\cr &&     \qquad \qquad
 \{ \widehat{\theta}_{\alpha}, \widehat{\theta}_{\beta} \}
= 0
\end{eqnarray}
and the primitive   coproducts:
\begin{eqnarray}
\label{luk-luk12yyb}
  \Delta \widehat{z}_{\mu} = \widehat{z}_{\mu}
\otimes 1 + 1 \otimes \widehat{z}_{\mu}
  \qquad
  \Delta \widehat{\theta}_{\alpha} =
\widehat{\theta}_{\alpha} \otimes 1 + 1 \otimes
  \widehat{\theta}_{\alpha}
\end{eqnarray}
The $\kappa$--deformed chiral superfield theory is obtained by
cosidering suitably ordered superfields. In the following Section
we shall consider the superFourier transform of deformed
superfields and consider the $\kappa$--deformed chiral superfield
theory.

\section{Fourier Supertransforms and $\kappa$--deformed Berezin
Integration}
\setcounter{equation}{0}

i) Fourier supertransform on classical superspace.

The superfields are defined as functions on superspace. Here we shall
restrict ourselves to $D=4$ chiral superspace
$z_A=(z_{\mu},\theta_{\alpha})$  $(\mu=0,1,2,3; \alpha=1,2)$ and
to chiral superfields $\Phi(z,\theta)$.

The Fourier supertransform
of the chiral superfield and its inverse take the form:

 \renewcommand{\theequation}{\thesection.1\alph{equation}}
 \setcounter{equation}{0}
\begin{eqnarray}
\label{luk-trzjea}
  \Phi(x,\theta) = {1\over (2\pi)^{2}}  \int d^4 p \, d^2 \eta\,
  \widetilde{\Phi}(p,\eta) e^{i(px+\eta\theta)}
\end{eqnarray}

\begin{eqnarray}
\label{luk-trzjeb}
  \widetilde{\Phi}(p,\eta) = {1\over (2\pi)^{2}}  \int d^4 x \, d^2
  \theta\,
  \Phi(x,\theta) e^{-i(px+\eta\theta)}
\end{eqnarray}
The Fourier supertransforms were considered firstly in
\cite{luk-lu26,luk-lu27}. It appears that the set of even and odd
variables ($z_{\mu},\theta_{\alpha}; p_{\mu},\eta_{\alpha}$)
describes the superphase space, with Grassmann variables
$\eta_{\alpha}$ describing ``odd momenta". The Berezin integration
rules    are valid in both odd position and momentum sectors:
 \renewcommand{\theequation}{\thesection.2\alph{equation}}
 \setcounter{equation}{0}
\begin{eqnarray}
\label{luk-trzdwa}
  \int d^2 \theta = \int  d^2 \theta\, \theta_{\alpha} = 0
  \qquad
  {1\over 2} \int  d^2 \theta \, \theta_{\alpha}\, \theta^{\alpha} =
  1
\end{eqnarray}

\begin{eqnarray}
\label{luk-trzdwab}
  \int d^2 \eta = \int  d^2 \eta \,\eta_{\alpha} = 0
  \qquad
  {1\over 2} \int  d^2 \theta\, \eta_{\alpha} \, \eta^{\alpha} =
  1
\end{eqnarray}
where $\eta^{\alpha}=\epsilon^{\alpha\beta}\eta_{\beta}$
and
 $\eta_{\alpha}\eta^{\alpha}=2\eta_{1}\eta_{2}$. It is   easy to
 see that $\theta^{2}={1\over 2} \theta_{\alpha}\theta^{\alpha}$
 $\eta^{2}={1\over 2} \eta_{\alpha}\eta^{\alpha}$ play the role of
 Dirac deltas, because
 \renewcommand{\theequation}{\thesection.3\alph{equation}}
 \setcounter{equation}{0}
\begin{eqnarray}
\label{luk-trztrza}
  \int d^2 \theta \, \theta^2 \, \Phi(z,\theta) =
    \Phi(z,\theta) \mid_{\theta=0}
    \end{eqnarray}

\begin{eqnarray}
\label{luk-trztrzb}
  \int d^2 \eta \, \eta^2\, \widetilde{\Phi}(p,\eta) =
    \widetilde{\Phi}(p,\eta) \mid_{\eta=0}
    \end{eqnarray}

 \renewcommand{\theequation}{\thesection.4\alph{equation}}
 \setcounter{equation}{0}
    The  formulae
 (\ref{luk-trzjea})--(\ref{luk-trzjeb}) in   component formalism
\begin{eqnarray}
\label{luk-trztrztrz}
   \Phi(z,\theta)= \Phi(z)+\Psi^{\alpha}(z)\theta_{\alpha}
  + F(z)\theta^{2}
\end{eqnarray}
lead to
\begin{eqnarray}
\label{luk-trzyy}
   \widetilde{\Phi}(p,\eta)=
   \widetilde{F}(p)- \widetilde{\Psi}^{\nu}(p)\eta_{\nu}
   - \widetilde{\Phi}(p)\eta^{2}
\end{eqnarray}
Let us consider  for example
 the chiral vertex $\Phi^{3}(z,\theta)$, present in
Wess--Zumino model. This vertex can be written in momentum
superspace as follows:
\renewcommand{\theequation}{\thesection.\arabic{equation}}
\setcounter{equation}{4}
\begin{eqnarray}
\label{luk-trzpi}
&&\lefteqn{\int d^4 z  \, d^2 \theta \, \Phi^3 (z,\theta) =
\int d^4 p_1 \ldots  d^4 p_3 \, d^2 \eta_1 \ldots d^2 \eta_3}
\cr
&& \quad \cdot \Phi(p_1,\eta_1)\,   \Phi(p_2,\eta_2)\,
 \Phi(p_3,\eta_3) \delta^4 (p_1 +p_2 +p_3)
 (\eta_1 + \eta_2 +\eta_3 )^2
\end{eqnarray}
We see therefore that in Feynmann  superdiagrams the chiral vertex
 (\ref{luk-trzpi}) will be represented by the product of Dirac deltas
describing the conservation at the vertex of the fourmomenta as well
as the Grassmann odd momenta.

ii) Fourier supertransform on  $\kappa$--deformed superspace.

Following the formulae
 (\ref{luk-dwanaxxa})--(\ref{luk-luk12yyb})
 we obtain the supersymmetric
extension of of $\kappa$--deformed Minkowski space to $\kappa$--deformed
superspace  $\widehat{x}_{\mu} \longrightarrow
 (\widehat{x}_{\mu},\widehat{\theta}_{\alpha})$. The ordered
 superexponential is defined as follows:
\begin{eqnarray}
\label{luk-trzsz}
  :e^{i(p_{\mu}\widehat{z}^{\mu} + \eta_{\alpha}\widehat{\theta}^{\alpha})}:
=  e^{-ip_{0}\widehat{z}_{0}}\,
e^{i(\vec{p}\vec{z} + \eta^{\alpha}\widehat{\theta}_{\alpha})}
\end{eqnarray}
where $(p_{\mu},\theta_{\alpha})$ satisfy the Abelian graded
algebra
 )\ref{luk-dwtrz}), i.e.
\begin{eqnarray}
\label{luk-trzsie}
  [p_{\mu},p_{\nu}] =[p_{\mu},\eta_{\alpha}]
  =\{ \eta_{\alpha},\eta_{\beta}\} = 0
\end{eqnarray}
From the formulae
 (\ref{luk-dwdw})--(\ref{luk-dwtrz})
 and
 (\ref{luk-trzsz})--(\ref{luk-trzsie}) follows that:
\begin{eqnarray}
\label{luk-trzos}
  :e^{i(p_{\mu}\widehat{z}^{\mu} + \eta_{\alpha}\widehat{\theta}^{\alpha})} : \,
  :e^{i(p^{\prime}_{\mu}\widehat{z}^{\mu} +
  \eta^{\prime}_{\alpha}\widehat{\theta}^{\alpha})}:= :
  e^{i\Delta^{(2)}_{\mu}(p,p^{\prime})\widehat{z}^{\mu}+
  \Delta^{(2)}_{\alpha}(\eta,\eta^{\prime})\widehat{\theta}^{\alpha}}:
\end{eqnarray}
where
\begin{eqnarray}
\label{luk-trzdzie}
\Delta_{0}(p,p^{\prime} ) &=& p_{0}+p_{0}^{\prime}
\cr
\Delta_{i}(p,p^{\prime} ) &=& p_{i}+ e^{-  {p_{0}\over \kappa}}
p^{\prime}_{i}
\cr
\Delta_{\alpha}(\eta,\eta^{\prime} )& =&
\eta_{\alpha}+ e^{-  {p_{0}\over  2\kappa}}
\eta^{\prime}_{\alpha}
\end{eqnarray}
The $\kappa$--deformed Fourier supertransform can be defined as
follows:
\begin{eqnarray}
\label{luk-trzdzies}
  \Phi(\widehat{z},\widehat{\theta}): =
  {1\over (2\pi)^{2}}  \int d^4 p \, d^2 \eta  \,
  \widetilde{\Phi}_{\kappa}(p,\eta):
  e^{i(p \widehat{z} +       \eta\widehat{\theta})}:
\end{eqnarray}
If we define inverse Fourier supertransform
\begin{eqnarray}
\label{luk-trzydzie}
  \widehat{\Phi}(p,\eta) =
  {1\over (2\pi)^{2}}
   \int d^4 \widehat{z} \,
   d^2 \widehat{\theta}  \,
   \Phi (\widehat{z},\widehat{\theta })
:  e^{- i(p \widehat{z}+
\eta\widehat{\theta})}:
\end{eqnarray}
under the   assumption that $(\widehat{\theta}^{2} = {1\over 2}
 \widehat{\theta}_{\alpha} \widehat{\theta}^{\alpha} )$
\renewcommand{\theequation}{\thesection.12\alph{equation}}
\setcounter{equation}{0}
\begin{eqnarray}
\int   d^2  \, \widetilde{\theta} \, \widetilde{\theta}^{2} = 1
\end{eqnarray}
or equivalently
($\eta^{2} \equiv {1\over 2} \eta_{\alpha}\eta^{\alpha}$)

\begin{eqnarray}
\label{luk-trzdwan}
   {1\over (2\pi)^{4}}  \iint d^4 \widehat{z} \,
   d^2 \widehat{\theta}:  \,
 e^{i(p \widehat{z}+
\eta\widehat{\theta})}: =
\delta^4 (p) \, \cdot \,  \eta^2
\end{eqnarray}
\renewcommand{\theequation}{\thesection.\arabic{equation}}
\setcounter{equation}{12}
one gets
\begin{eqnarray}
\label{luk-trztrzyn}
 \widehat{\Phi}_{\kappa}({p},{\eta}) =
 e^{- {4 p_{0}\over  \kappa}}
 \widetilde{\Phi}
 \left(  e^{ {p_{0}\over  \kappa}} \vec{p}, p_0 ,
  e^{  {p_{0}\over  2\kappa}} \eta_{\alpha} \right)
  \end{eqnarray}

  For $\kappa$--deformed chiral fields one can consider their
  local powers, and perform the $\kappa$--deformed superspace
  integrals. One gets

 \renewcommand{\theequation}{\thesection.14\alph{equation}}
\setcounter{equation}{0}

\begin{eqnarray}
\label{luk-trzcztera}
&&  \iint d^4 \widehat{z} \, d^2  \widehat{\theta} :
  \Phi(\widehat{z},\widehat{\theta})
  = \widehat{\Phi}(0,0)
  \cr\cr
  && \iint d^4 \widehat{z} \, d^2  \widehat{\theta}
  \Phi^{2}(\widehat{z},\widehat{\theta})
  = \int d^4 p_1 \, d^4 p_2 \, d^2   \eta_1 \, d^2 \eta_2 \,
 \\ \cr
&&  \widetilde{\Phi}_{\kappa}(p_1, \eta_1 )\,
  \widetilde{\Phi}_{\kappa}(p_2, \eta_2 )\,
\delta(p_{01} + p_{02} )\delta^{(3)}
\left(
\vec{p}_{1} + e^{p_{01}\over \kappa} \, \vec{p}_{2} \right)
  (\eta_1 +
  e^{ {p_{01}\over  2\kappa}} \eta_2 )^2
\cr\cr
&&\iint d^4 \widehat{z} \, d^2  \widehat{\theta}
  \Phi^{3}(\widehat{z},\widehat{\theta})=
  \int \prod\limits^{3}_{i=1} d^4 p_i \,
  d^2 \eta_{i} \, \cdot
  \, \widetilde{\Phi}_{\kappa} (p_{i},\eta_{i})
  \cr
  &&  \qquad \cdot\  \delta(p_{01} + p_{02} +  p_{03} ) \cdot
 \delta^{(3)}
\left(
       \vec{p}_{1} + e^{p_{01}\over \kappa} \, \vec{p}_{2}
+ {p_{0} + p_{02} \over \kappa} \, \vec{p}_{3} \right)
\cr
&& \qquad
\cdot \
\left(\eta_{1} +
 e^{ {p_{02}\over  2\kappa} \eta_2}
 +
  e^{ {p_{01} +    p_{02} \over  \kappa} \eta_3}\right)^2
\label{luk-trzczterb}
\end{eqnarray}
The formulae
 (\ref{luk-trzcztera})--(\ref{luk-trzczterb}) can 
 be used for the description of
$\kappa$--deformed vertices in Wess--Zumino model for chiral superfields.

\section{Star Product for $\kappa$--deformed Superfield Theory}
\setcounter{equation}{0}

In this section we shall extend the star product for the functions
on $\kappa$--deformed Minkowski space given in
\cite{luk-kos3}
  to the case of
functions on   $\kappa$--deformed chiral superspace, described by
the relations
 (\ref{luk-dwanaxxa})--(\ref{luk-luk12yyb}).

The CBH $\star$--product formula for unordered
 exponentials takes the form

\renewcommand{\theequation}{\thesection.\arabic{equation}}
\setcounter{equation}{0}                                   

\begin{eqnarray}
\label{luk-cztejed}
 e^{ip_{\mu}z^{\mu} +\overline{\eta}_{\dot{\alpha}}
  \overline{\theta}^{\dot{\alpha}} } \cdot
  e^{ip_{\nu}^{\prime} z^{\nu}
  +\overline{\eta}_{\dot{\beta}}^{\prime}
  \overline{\theta}^{\dot{\beta}}}
 =
  e^{i\gamma_{\mu}(p,p^{\prime})z^{\mu}   
   +\overline{\sigma}_{\dot{\alpha}}(p,p^{\prime},
  \overline{\eta},\overline{\eta}^{\prime})
   \overline{\theta}^{\dot{\alpha}} }
\end{eqnarray}
where
\renewcommand{\theequation}{\thesection.2\alph{equation}}
\setcounter{equation}{0}
\begin{eqnarray}
\label{luk-czterdw}
  \gamma_{0} &= & p_{0} + p^{\prime}_{0}
\\
\gamma_{k} & = & {p_{k} \, e^{ {p^{\prime}_{0}\over \kappa}}
f\left( {
p_{0}\over \kappa } \right) + p^{\prime}_{k} f \left(
{p^{\prime}_{0}\over \kappa } \right)
\over
f \left( {P_{0}+ p^{\prime}_{0} \over \kappa}\right) }
\\
\overline{\sigma}_{\dot{\alpha}}
& = & {
\overline{\eta}_{\dot{\alpha}} \,
 e^{ {p^{\prime}_{0}\over 2 \kappa}}
f\left({p_{0}\over 2\kappa } \right)
+ \overline{\eta}^{\prime}_{\dot{\alpha}}
f \left(
{ p^{\prime}_{0}\over 2 \kappa } \right)
\over
f \left( {P_{0}+ p^{\prime}_{0} \over 2 \kappa}\right) }
\end{eqnarray}
and $f(x)\equiv {e^{x} - 1   \over x}$. The star  product
multiplication reproduces the formula 
 (\ref{luk-cztejed}).
\renewcommand{\theequation}{\thesection.\arabic{equation}}
\setcounter{equation}{2}
\begin{eqnarray}
\label{luk-cztedwaa}
 e^{ip_{\mu}z^{\mu} +\overline{\eta}_{\dot{\alpha}}
  \overline{\theta}^{\dot{\alpha}} } \star
  e^{ip_{\nu}^{\prime} z^{\nu}
  +\overline{\eta}_{\dot{\beta}}^{\prime}
  \overline{\theta}^{\dot{\beta}}}
 =
  e^{i\gamma_{\mu}(p,p^{\prime})z^{\mu} 
   +\overline{\sigma}_{\dot{\alpha}}(p,p^{\prime},
  \overline{\eta},\overline{\eta}^{\prime})
   \overline{\theta}^{\dot{\alpha}}      }
\end{eqnarray}
For arbitrary superfields $\phi(z,\theta)$ and
$\chi(z,\theta)$ one gets
\begin{eqnarray}
\label{luk-czterczter}
\lefteqn{\phi(z,\theta)      \star \chi(z,\theta) = }
\cr\cr
&&\hspace{-1truecm} \ = \phi \left(
{1\over i} \, {\partial \over \partial p_{\mu}}
, {\partial \over \partial \overline{\eta}_{\dot{\alpha}} }
\right)
\chi \left(
{1 \over i}  \, {\partial \over \partial p_{\mu}^{\prime}}
, {\partial \over \partial \overline{\eta}_{\dot{\alpha}}^{\prime} }
\right)
e^{i\gamma_{\mu}(p,p^{\prime}) z^{\mu}
+ \overline{\sigma}_{\dot{\alpha}}
(p,p^{\prime},\overline{\eta},\overline{\eta}^{\prime} )
\overline{\theta}^{\dot{\alpha}} }\left|_{
\begin{subarray}{l}
p=0
\cr
p^{\prime}=0
\cr
 \overline{\eta}=0
\cr
\overline{\eta}^{\prime} = 0
\end{subarray}
 }
\right.
\end{eqnarray}
or equivalently

\begin{eqnarray}
\label{luk-czterpiec}
\phi(z,\theta)      \star \chi(z,\theta) &= &
e^{i z^{\mu}
\left(\gamma_{\mu}
\left(
{\partial \over \partial y},
 {\partial \over \partial y^{\prime} }
\right)
 -
 {\partial \over \partial y^{\mu}}
-  {\partial \over \partial y^{\prime \mu} }
\right)
- \overline{\theta}^{\dot{\alpha}}
\left(
\overline{\sigma}^{\dot{\alpha}}
\left(
{\partial \over \partial y}
, {\partial \over \partial y^{\prime} },
{\partial \over \partial \omega}
, {\partial \over \partial \omega^{\prime} }
\right)
-
{ \partial \over \partial \omega_{\dot{\alpha}} }
- {\partial \over \partial \omega^{\prime}_{\dot{\alpha} } }
\right)}
\cr
&& \cdot \phi(y,\omega) \chi(y^{\prime},\omega^{\prime})
\left|_{
\begin{subarray}{l}
y= y^{\prime}=z
\cr 
 \omega=\omega^{\prime}=\overline{\theta}
\end{subarray}
 }
\right.
\end{eqnarray}
In particular we get

\begin{eqnarray}
\label{luk-czterszes}
z^{i} \star z^{j} & = & z^{i} z^{j}
\cr
z^{0} \star z^{i} & = & z^{0} z^{i}
 + {i\over 2 \kappa} \, z^{i}
 \cr
z^{i} \star z^{0} & = & z^{0} z^{i}
 - {i\over 2 \kappa} \, z^{i}
 \cr
 z^{i} \star \overline{\theta}^{\dot{\alpha}}
 & = & z^{i} \overline{\theta}^{\dot{\alpha}}
 \cr
\overline{\theta}^{\dot{\alpha}} \star z^{i}
 & = &  z^{i} \overline{\theta}^{\dot{\alpha}}
 \cr
 z^{0} \star \overline{\theta}^{\dot{\alpha}}
 & = & z^{0} \overline{\theta}^{\dot{\alpha}}
+
  {i\over 4 \kappa}
  \overline{\theta}^{\dot{\alpha}}
  \cr
    \overline{\theta}^{\dot{\alpha}} \star z^{0}
 & = & z^{0} \overline{\theta}^{\dot{\alpha}}
-
 {i\over 4 \kappa}
  \overline{\theta}^{\dot{\alpha}}
  \cr
    \overline{\theta}^{\dot{\alpha}}\star
      \overline{\theta}^{\dot{\beta}}
      &= &  \overline{\theta}^{\dot{\alpha}}
        \overline{\theta}^{\dot{\beta}}
\end{eqnarray}
Star product $\circledast$ corresponding to the multiplication of
ordered exponentials
 (\ref{luk-trzsz}) takes the form:

\begin{eqnarray}
\label{luk-cztersie}
 &&
 e^{ip_{\mu}z^{\mu}
 + \overline{\eta}_{\dot{\alpha}}
  \overline{\theta}^{\dot{\alpha}}
  }
 \circledast
  e^{i p_{\mu}^{\prime} z^{\mu}
  +\overline{\eta}_{\dot{\alpha}}^{\prime}
  \overline{\theta}^{\dot{\alpha}}
}
\cr
\cr
&& =  e^{i (p_{0} + p^{\prime}_{0} z^{0} 
 + i
(
 e^ { p^{\prime}_{0} \over \kappa  }
  p_{\kappa}
 + p^{\prime}_{\kappa} 
)
 z^{\kappa}
+
(
 e^{ p^{\prime}_{0} \over 2\kappa }
  \overline{\eta}_{\dot{\alpha }}
 + \overline{\eta}_{\dot{\alpha}}^{\prime}
)
 \overline{\theta}^{\dot{\alpha}}    }
 \end{eqnarray}

The superalgebra
 (\ref{luk-dwanaxxa})--(\ref{luk-luk12yyb})  of $\kappa$--deformed superspace is
obtained from the following relations:

\begin{eqnarray}
\label{luk-czterosi}
z^{k} \circledast   \overline{\theta}^{\dot{\alpha}} &=&
\overline{\theta}^{\dot{\alpha}}
\circledast z^{k} =  z^{k}
\overline{\theta}^{\dot{\alpha}}
\cr
\overline{\theta}^{\dot{\alpha}}
\circledast
\overline{\theta}^{\dot{\beta}} & = &
\overline{\theta}^{\dot{\alpha}}
\overline{\theta}^{\dot{\beta}}
 \cr
z^{k} \circledast z^{i} & = & z^{k} z^{i}
 \cr
 z^{0} \circledast z^{i} & = & z^{0} z^{i}
 \cr
z^{i} \circledast z^{0}
 & = &  z^{0} z^{i} -   { i \over \kappa }
  z^{i}
 \cr
 z^{0} \circledast \overline{\theta}^{\dot{\alpha}}
 & = & z^{0} \overline{\theta}^{\dot{\alpha}}
  \cr
    \overline{\theta}^{\dot{\alpha}} \circledast z^{0}
 & = & z^{0} \overline{\theta}^{\dot{\alpha}}
-
 {i\over 2\kappa}
  \overline{\theta}^{\dot{\alpha}}
\end{eqnarray}

Similarly  like in nonsupersymmetric case the star--product 
 (\ref{luk-cztersie})
is more physical because reproduces the composition law of even
and odd momenta consistent with coalgebra structure.

\section{Final Remarks}
\setcounter{equation}{0}

In this lecture      we outlined present status of quantum
deformations of space--time supersymmetries\footnote{We did not
consider here however, the quantum deformations of infinite --
parameter superconformal symmetries in $1+1$  dimensions,
described by superVirasoro algebras as well as affine
$OSp(N;2)$--superalgebras}, and for the case of
$\kappa$--deformation of $D=4$
supersymmetries proposed the corresponding deformation of chiral
 superfield theory. It appears that only the $\kappa$--deformed chiral
 superspace generators describe a closed subalgebra
 of $\kappa$--deformed
 $D=4$
Poincar\'{e} group. At present it
 can be obtained  the $\kappa$--deformation of
superfield theory on real superspace  can be obtained. The deformation of chiral
superfield theory can be described by the    following diagram:

\begin{figure}[htb]
  \centering
  {\epsfig{file=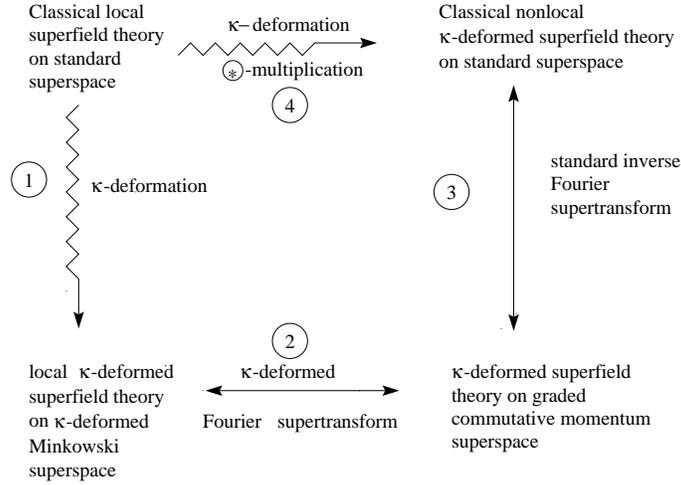,width=9cm}}
  \caption{$\kappa$--deformation of local  superfield theory}
\label{luk-fig1}
\end{figure}

The star product $\circledast $ given by formula 
 (\ref{luk-cztersie}) (see \cztop  on Fig. 1)
  is selected by the  choice of superFourier transform
 (\ref{luk-trzdzies}), with ordered
 Fourier exponential described by
 (\ref{luk-trzsz}). Equivalently, the
$\circledast $--product multiplication can be obtained by the following
three consecutive steps:

i) Deformation of local superfield theory (see
\jedop  on Fig. 1)

ii) $\kappa$--deformed superfield transform
 (\ref{luk-trzdzies}) (see \dwaop on Fig. 1)

iii) inverse classical Fourier transform (see \trzop on Fig. 1)

\begin{eqnarray}
\label{luk-piejede}
  \Phi(z,\theta) = {1\over (2\pi)^{2} }
  \int d^{4} p \, d^{2} \theta \,
  e^{-i (p_{\mu}z^{\mu} + \eta_{\alpha}\theta^{\alpha})}
  \widetilde{\Phi}(p,\eta)
\end{eqnarray}
obtained in the limit $\kappa \to \infty$ from the inverse
Fourier transform
 (\ref{luk-trzydzie}).

Finally it should be observed that for the deformation 
 (\ref{luk-lukjedcc}) 
with constant $\widehat{\theta}_{\mu\nu}$ there were calculated
some explicite corrections to physical processes, in particular
for $D=4$ QED
 \cite{luk-lu26}--\cite{luk-lu28}.
We   would like to   stress
  that these
 calculations should  be repeated for Lie algebraic deformations
 of space--time and superspace, in particular in the
 $\kappa$--deformed framework. The preliminary results in this
 direction has been obtained in \cite{luk-lu32,luk-lu33}.



\begin{thebibliography}{99}

\bibitem{luk-dop} S. Dopplicher, K. Fredenhagen, J. Roberts, Phys.
Lett. {\b B331}, 39 (1994); Comm. Math. Phys. {\bf 172}, 187
(1995)

\bibitem{luk-gar} L.J. Garay, Int. Journ. Mod. Phys. {\bf A10}, 145
(1995)

\bibitem{luk-chon}Chong-Sun Chu, Pei-Ming Ho, hep--th/9812219;
hep--th/9906192

\bibitem{luk-seib} N. Seiberg, E. Witten, JHEP  9909: 032 (1999)

\bibitem{luk-mad} J. Madore, S. Schraml, P. Schupp, J. Wess,
 hep---th/0001203

\bibitem{luk-chu} C. Chu, F. Zamorra, hep--th/9912153

\bibitem{luk-ferr} S. Ferrara, M. Lledo, hep--th/0002084

\bibitem{luk-tera} S. Terashima, hep--th/0002119

   \bibitem{luk-dop2} A.A. Bichl, J.M. Grimstrup, H. Grosse, L.
   Popp, M. Schweda, R. Wulkenhaar,  hep--th/0007050

\bibitem{luk-dopp} S. Dopplicher, Ann. Inst. Henri Poinc. {\bf 64}, 543 (1996)

\bibitem{luk-lukx} J. Lukierski, P. Stichel, W.J. Zakrzewski,
Ann,. Phys. {\bf 261}, 224 (1997)


   \bibitem{luk-luk1} J. Lukierski, A. Nowicki, J. Sobczyk,
   J. Phys. {\bf 26A}, L1109 (1993)

   \bibitem{luk-kos1} P. Kosi\'{n}ski, J. Lukierski, P. Ma\'{s}lanka,
    J. Sobczyk,
   J. Phys. {\bf 27A}, 6827 (1994);\newline
   ibid. {\bf 28A}, 2255 (1995)

   \bibitem{luk-kos2}P. Kosi\'{n}ski, J. Lukierski, P. Ma\'{s}lanka,
     J. Sobczyk,
   J. Math. Phys. {\bf 37},3041 (1996)

   \bibitem{luk-kos3}P. Kosi\'{n}ski, J. Lukierski, P. Ma\'{s}lanka,
   hep--th/0009120

   \bibitem{luk-chai} M. Chaichian, P.P. Kulish, Phys. Lett. {\bf
   B234}, 72 (199)

   \bibitem{luk-khoro} S.M. Khoroshkin, V.N. Tolstoy, Comm. Math.
   Phys. {\bf 141}, 599 (1991)

   \bibitem{luk-luk3} J. Lukierski, A. Nowicki, Phys. Lett.
   {B279}, 299 (1992)

   \bibitem{luk-dobr} V. Dobrev, Journ. Phys. {\bf A26}, 1317
   (1993)

   \bibitem{luk-??2}L. Dabrawski, V.K. Dobrev, R. Floreanini, V.
   Husain, Phys. Lett. {\bf B302}, 215 (1993)

   \bibitem{luk-??3}   C. Juszczak, Journ. Phys. {\bf A27}, 385
   (1994)

   \bibitem{luk-luk4} J. Lukierski, P. Minnaert, M. Mozrzymas, Phys.
   Lett. {\bf B371}, 215 (1996)

   \bibitem{luk-jusz} C. Juszczak,  J. Sobczyk,
   math.QA/9809006

   \bibitem{luk-cele} R. Celeghini, P. Kulish,  q--alg/9712006,

   \bibitem{luk-herr} F. Herranz, Journ. Phys. {\bf A30}, 6123
   (1997)

   \bibitem{luk-moy} S. Majid, ``Foundations of Quantum Group Theory"
      Cambridge Univ. Press, 1995

\bibitem{luk-lu2525} D. Leites, B.M. Zupnik ``Multiple Processes at High
 Energies" (in Russian), Tashkent 1976, p. 3--26

\bibitem{luk-berez} F.A. Berezin, M.S. Marinov, Ann. Phys. {\bf 104},
336 (1977)

\bibitem{luk-lu26} I. Mocioiu, M. Pospelov, R. Roiban,
  Phys. Lett. {\bf B489}, 390 (2000)

\bibitem{luk-lu27} N. Chair, M.M. Sheikh--Jabbari, hep--th/0009037

\bibitem{luk-lu28} M. Chaichian, M.M. Sheikh--Jabbari,
 A. Tureanu hep--th/0010175


\bibitem{luk-lu32} J. Lukierski, H. Ruegg, W. Ruehl, Phys.
 Lett. {\bf 313}, 357 (1993)

\bibitem{luk-lu33} L.C. Biedenharn, B. Mueller, M. Tarlini,
Phys. Lett. {\bf 318}, 613 (1993)

 \end{thebibliography}
\end{document}